\documentclass[aps,prl,twocolumn,groupedaddress,floatfix,showpacs]{revtex4-1}
\usepackage{amssymb,amsmath}
\usepackage{graphicx}
\usepackage{subfigure}
\usepackage[english]{babel}
\usepackage{float}
\usepackage{color}

\begin{document}
\newcommand{\be}{\begin{equation}}
\newcommand{\ee}{\end{equation}}
\newcommand{\rojo}[1]{\textcolor{red}{#1}}

\title{Solitons in a modified discrete nonlinear Schr\"{o}dinger equation}

\author{Mario I. Molina}

\address{Departamento de F\'{\i}sica and MSI-Nucleus on Advanced Optics, Facultad de Ciencias, Universidad de Chile, Casilla 653, Santiago, Chile}


\begin{abstract}

We study the bulk and surface nonlinear modes of the modified one-dimensional discrete nonlinear Schr\"{o}dinger (mDNLS) equation. A linear and a modulational stability analysis of the lowest-order modes is carried out. While for the fundamental bulk mode there is no power threshold, the fundamental surface mode needs a minimum power level to exist. 
Examination of the time evolution of  discrete solitons in the limit of strongly localized modes, suggests ways to manage the Peierls-Nabarro barrier, facilitating in this way a degree of steering. 
The long-time propagation of an initially localized excitation shows that, at long evolution times, nonlinear effects become negligible and as a result, the propagation becomes ballistic. The similarity of all these results to the ones obtained for the DNLS equation, points out to the robustness of the discrete soliton phenomenology.

\end{abstract}

\maketitle
\section{Introduction}

In the semiclassical approach to the coupled electron-phonon problem, the electronic degrees of freedom are coupled to the vibrational ones, where the latter are pictured as classical oscillators. A further approximation assumes that these oscillators are enslaved to the electron, thus reducing the number of equations which now contain only electronic coordinates. When the oscillators are pictured as Einstein or, optical oscillators, one arrives to an effective electronic equation known as the Discrete Nonlinear (DNLS) equation\cite{davydov}:
\begin{eqnarray}
i {dC_{n}\over{dt}} + V (C_{n+1} + C_{n-1}) + \chi |C_{n}|^2 C_{n}=0.\label{one}
\end{eqnarray}
Here, $C_{n}$ is the electronic probability amplitude at site $n$, $V$ is the coupling between nearest neighbor sites, and $\chi$ is the nonlinearity parameter, proportional to the square of the electron-phonon coupling. One of the main consequences of the well-studied DNLS equation is the existence of a long-lived nonlinear excitation termed ``discrete soliton''. This type of excitation is believed to be of great importance for the trapping and transmission of energy in biomolecules\cite{biomolecules}.

On the other hand, when the oscillators are taken as of the Debye, or acoustic type, one arrives to the less-known Modified Discrete Nonlinear (mDNLS) equation:
\begin{eqnarray}
i {dC_{n}\over{dt}} + V (C_{n+1} + C_{n-1}) &+&\nonumber\\
 \chi (|C_{n+1}|^2 + |C_{n-1}|^2 + 2 |C_{n}|^2) C_{n}&=&0.\label{two}
\end{eqnarray}

In this work we examine the selftrapping and transport properties  of the mDNLS equation (\ref{two}), and show
that its soliton phenomenology is similar to the one found previously in DNLS. This is very important since it supports the idea of a discrete soliton as a robust
excitation of the system, regardless of the precise nature of the underlying phonons. In this sense, the DNLS and mDNLS can be regarded as complementary equations.

Equation (\ref{two}) has two conserved quantities: The power $P=\sum_{n} |C_{n}|^2$, and the Hamiltonian 

\begin{eqnarray}
H &=& V \sum_{n} (C_{n} C_{n+1}^{*} + C_{n}^{*} C_{n+1})\nonumber\\& & + \chi \sum_{n} ( |C_{n+1}|^2 |C_{n+2}|^2+|C_{n}|^4). \label{H}
\end{eqnarray}
Our system is a Hamiltonian one since from Eq.(\ref{H}) one obtains Eqs.(\ref{one}) by using the canonical equations
$dq_{n}/dt = \partial H/\partial p_{n}$, $dp_{n}/dt=-\partial H/\partial q_{n}$ with  canonically conjugate variables $q_{n}=C_{n}$ and $p_{n}= i C_{n}^{*}$. 
The mDNLS was first found in earlier studies of polaron formation, from the coupled electron-phonon equations in the adiabatic limit\cite{davydov,prb51,kopidakis}. For the case of a large polaron whose size is much larger than the lattice spacing, a further continuum approximation is possible and one arrives to the continuous Nonlinear Schr\"{o}dinger equation (NLS), which has a well-known soliton solution. 
The mDNLS has been used to study certain recurrences that occur in the coupled electron-phonon problem \cite{kopidakis,tsironis}. The selftrapping properties of the mDNLS were also examined in Ref.[\cite{tsironis}]. The DNLS and mDNLS equations are complementary and are useful to describe the dynamics of excitations.

In this work we will focus on the different families of nonlinear bulk and surface modes and their stability properties, transport exponents and on the propagation of mDNLS solitons, looking for possible means of propagation control. 

{\em Nonlinear modes}. The nonlinear modes are found by setting
$C_{n}(t)=\phi_{n} \exp{i \lambda t}$, which leads to
the nonlinear eigenvalue equation
\begin{eqnarray}
-\lambda \phi_{n}+ V(\phi_{n+1}+\phi_{n-1}) &+&\nonumber\\
\chi (|\phi_{n+1}|^2+ |\phi_{n-1}|^2 + 2 |\phi_{n}|^2) \phi_{n}&=&0.\label{eq:3}
\end{eqnarray}
For a given $\lambda$, the system of equations (\ref{eq:3}) is solved numerically by means of a multidimensional Newton-raphson scheme, using as a seed the form obtained from the decoupled limit, also known as the anticontinuous limit. Figures 1 and 2 show examples of the lowest-order nonlinear localized modes, for the bulk (away from $n=1$ or $n=N$) and the surface (near $n=1$ or $n=N$) respectively.
For the bulk case, we see an ``odd'' mode (A), and ``even'' mode (B) and two ``twisted'' modes (C and D). For the surface case we see a truncated ``odd'' mode (A), a ``flat top'' mode (B) and two ``twisted'' modes (C and D).
\begin{figure}[h]
\noindent
\includegraphics[scale=0.5,angle=0]{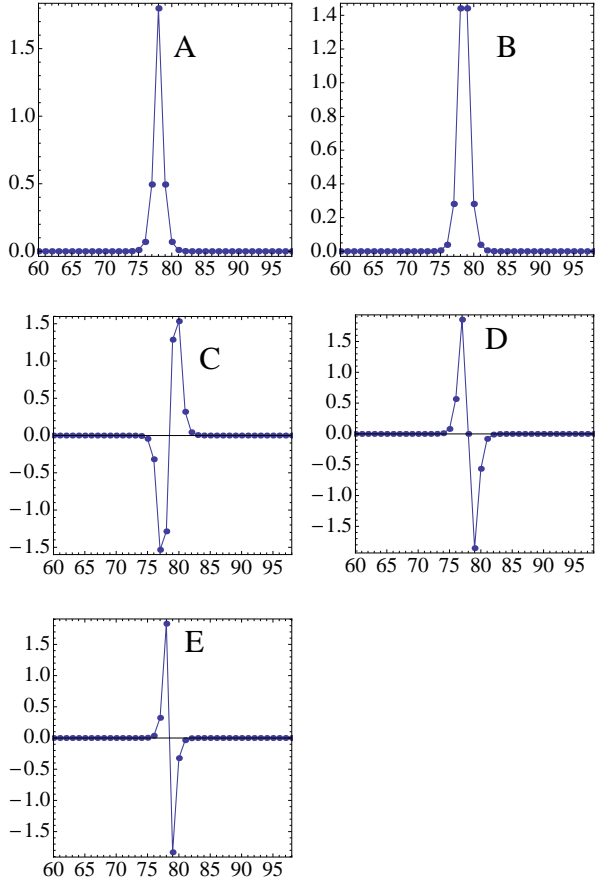}
\caption{Some lowest-order nonlinear bulk modes ($V=1, \lambda=9.3$). The modes shown here have different power content.}
\label{fig1}
\end{figure}
To compute the linear stability of the modes we introduce 
a weak perturbation as $C_{n}(t)=(\phi_{n} + \delta_{n}(t))\exp(i \lambda t)$, and obtain a linear evolution equation for $\delta_{n}(t)$, where $|\delta_{n}(t)|\ll|\phi_{n}|$. After decomposing $\delta_{n}(t) = x_{n}(t) + i y_{n}(t)$, and inserting into Eq.(\ref{one}), one arrives at a set of coupled real equations:
\be
{d\over{d t}}\vec{x} + {\bf A} \ \vec{y}=0, \ \ \ {d\over{d t}}\vec{y} + {\bf B} \ \vec{x}=0\label{eq:4}
\ee
\begin{figure}[t]
\noindent
\includegraphics[scale=0.5,angle=0]{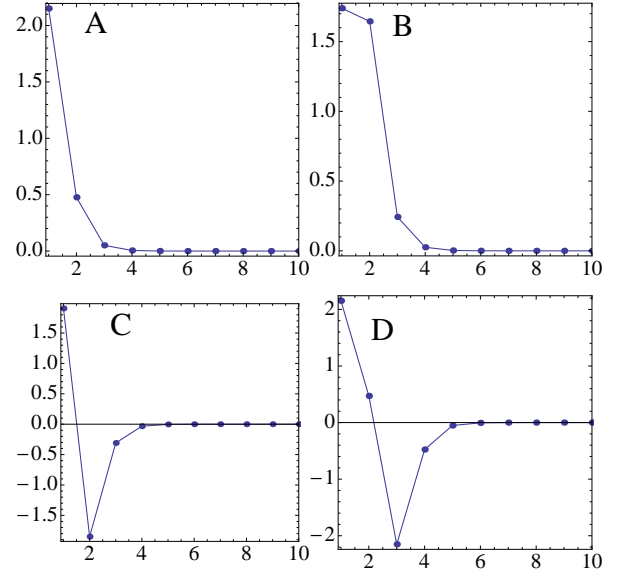}
\caption{Some lowest-order nonlinear surface modes ($V=1, \lambda=9.3$).  The modes shown here have different power content.}
\label{fig2}
\end{figure}
where {\cal x}=$(x_{1}, x_{2}, ...x_{N})$, 
{\cal y}=$(y_{1}, y_{2}, ...y_{N})$ and {\bf A} and {\bf B} are matrices defined by
\begin{eqnarray}
\lefteqn{\mbox{\bf A}_{n m}=}\nonumber \\
& & 
(-\lambda + \chi |\phi_{n+1}|^2+ \chi |\phi_{n-1}|^2+4 \chi |\phi_{n}|^2-2 \chi \phi_{n}^2) \delta_{n m}\nonumber\\
& & +(V + \chi \phi_{n} (\phi_{n+1}^{*}-\phi_{n+1})) \delta_{n,m-1} + \nonumber\\
& & (V + \chi \phi_{n} (\phi_{n-1}^{*}-\phi_{n-1})) \delta_{n,m+1}\label{eq5}
\end{eqnarray}
and
\begin{eqnarray}
\lefteqn{\mbox{\bf B}_{n m}=}\nonumber \\
& & 
(\lambda - \chi |\phi_{n+1}|^2 - \chi |\phi_{n-1}|^2-4 \chi |\phi_{n}|^2-2 \chi \phi_{n}^2) \delta_{n m}\nonumber\\
& & -(V + \chi \phi_{n} (\phi_{n+1}^{*}+\phi_{n+1})) \delta_{n,m-1} + \nonumber\\
& & -(V + \chi \phi_{n} (\phi_{n-1}^{*}+\phi_{n-1})) \delta_{n,m+1}\label{eq6}
\end{eqnarray}
From Eq. (\ref{eq:4}) one obtains
\be
{d^2\over{d t^2}}\vec{x} - {\bf A} {\bf B}\ \vec{x}=0, \ \ \ {d^2\over{d t^2}}\vec{y} - {\bf B} {\bf A} \ \vec{x}=0\label{eq:5}
\ee
Thus, the linear stability of the nonlinear modes is determined by the eigenvalue spectra of the matrices ${\bf A B}$ and ${\bf B A}$. A convenient parameter to quantify the stability of a mode is the instability gain G, defined as
\be
G = \mbox{Max of} \left\{ {1\over{2}} \left( Re[g] + \sqrt{Re[g]^2 + Im[g]^2}\  \right) \right\}^{1/2} 
\ee
over all $g$ values, where  $g$ is one of the eigenvalues of {\bf A B} (or {\bf B A}). When $G$ is zero, the mode is stable; otherwise it is unstable. This parameter is nothing else but the largest growth rate of the mode and is given by the imaginary part of the square root of the complex eigenvalue of {\bf A B} (or {\bf B A}).

\begin{figure}[h]
\noindent
\includegraphics[scale=0.8,angle=0]{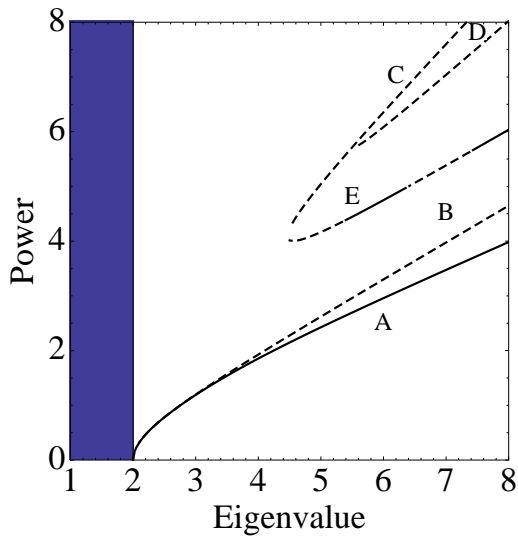}
\caption{Power content versus eigenvalue for the nonlinear modes of Fig.1. Continuous (dashed) curves denote stable (unstable) modes ($V=1, \chi=1$).}
\label{fig3}
\end{figure}
\begin{figure}[h]
\noindent
\includegraphics[scale=0.77,angle=0]{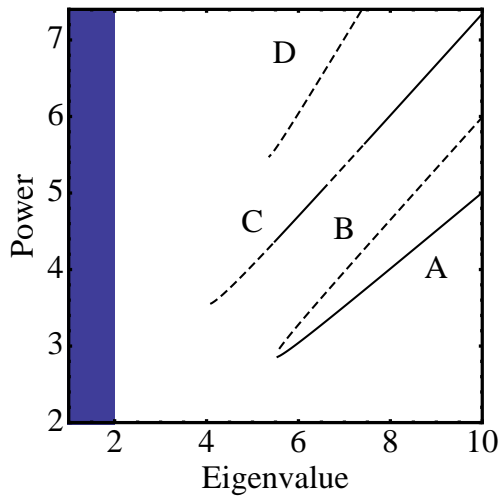}
\caption{Power content versus eigenvalue for the nonlinear modes of Fig.2. Continuous (dashed) curves denote stable (unstable) modes. ($V=1, \chi=1$)}
\label{fig4}
\end{figure}
Figures 3 and 4 show the power vs eigenvalue curves for some lowest-order modes, along with their stability. We note that, while for the bulk modes, there are at least two modes with no threshold power, for the surface modes they all require a minimum power threshold (nonlinearity) to exist. The only stable lowest-order bulk mode is the odd one, which is stable all the way down to the linear band. For the surface modes we observe that they all require a minimum power threshold to exist. Also, (not shown) we observe that all surface modes in Fig.2 can be easily continued below the ``surface  layer'' and, as the distance from the surface is increased, these modes converge to the bulk modes of. Fig.1

{\em Soliton propagation}
We consider here the propagation of an approximate soliton solution using Eq. (\ref{one}). As an initial condition we will use the form 
$u(0)= A\  \mbox{sech}[(A/\sqrt{2})(n-n_{c})]\ \exp[-i k (n - n_{c})]$ which is a discretization of the exact continuous NLS one-soliton solution. Parameter $k$ represents the initial momentum of the pulse and $n_{c}$ is the position of the soliton center. This ansatz is reasonable for wide solitons where the discrete character of the lattice is of no consequence. Figure 5 shows an example of discrete soliton propagation for two different values of momentum. In the case of large kick, the soliton propagates across the lattice and bounces elastically from the ends of the chain. For low values of $k$, the soliton propagates some short distance and gets selftrapped eventually around some lattice site. It should be noted that similar results are obtained for more generic 
\begin{figure}[h]
\noindent
\includegraphics[scale=0.225,angle=0]{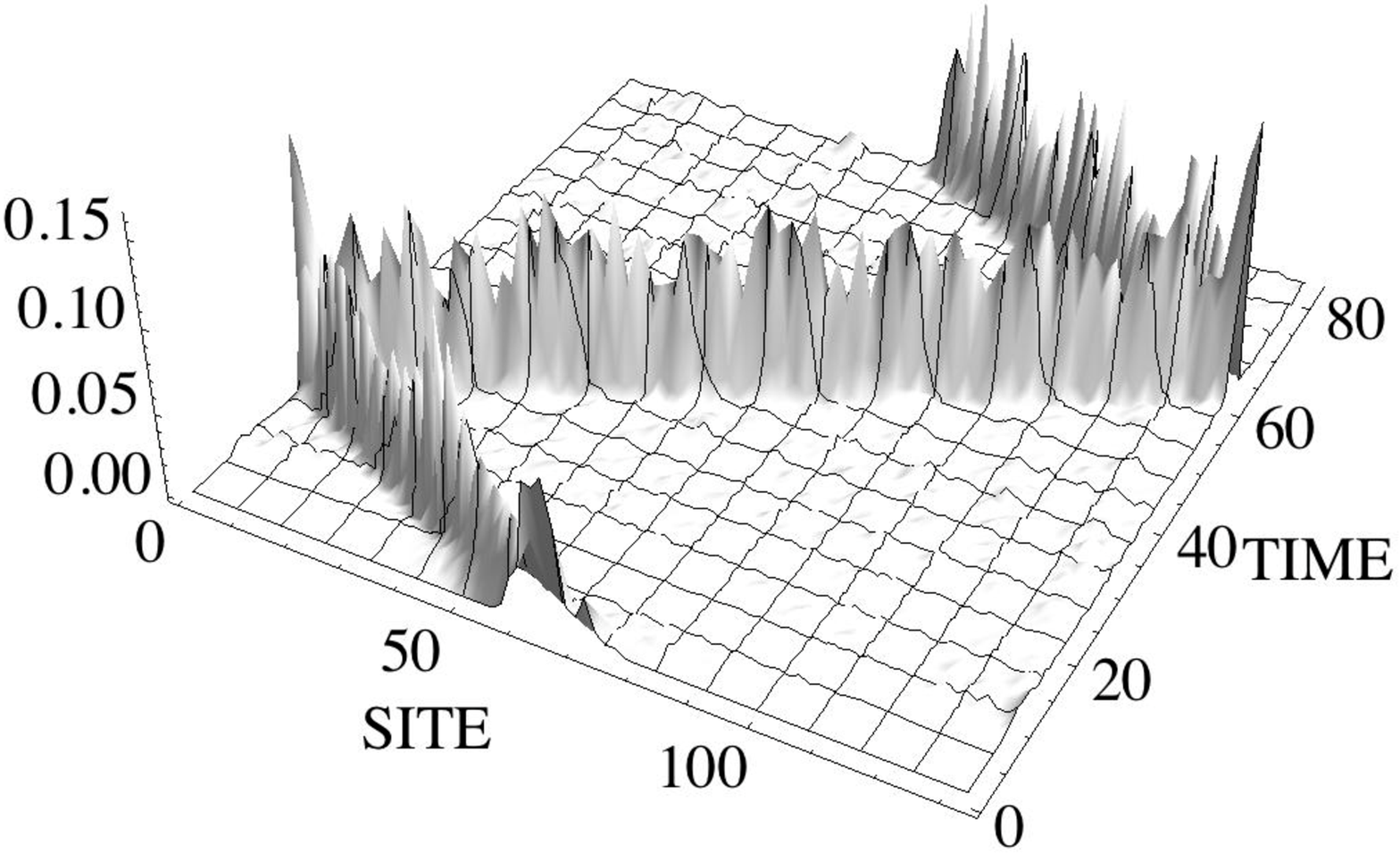}
\includegraphics[scale=0.225,angle=0]{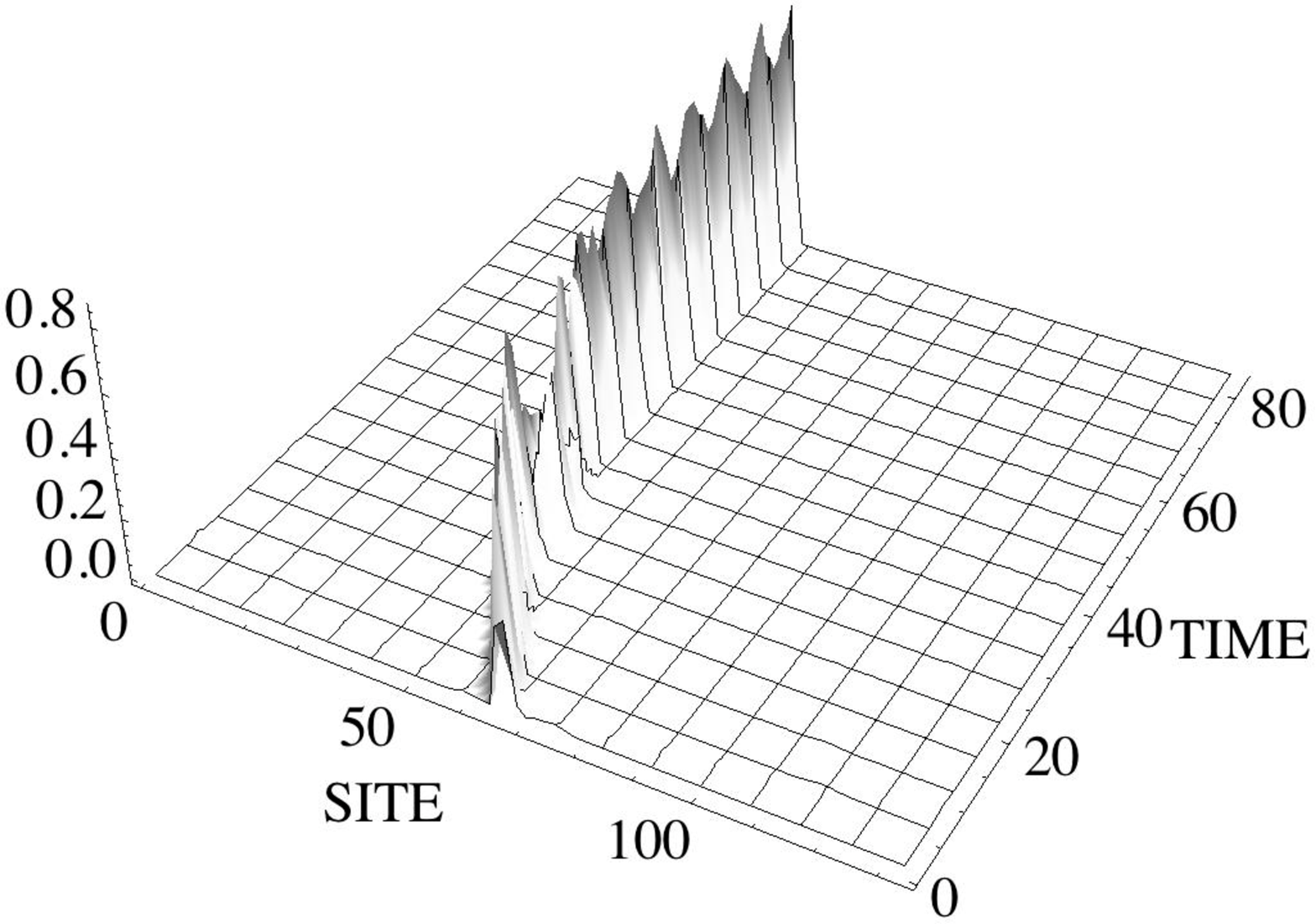}
\caption{Examples of propagation of a discrete mDNLS soliton for large and small values of the initial momentum. Top: $k=0.4, A=0.5$. Bottom: $k=0.2, A=0.8$. For both cases 
$V=1, \chi=1$.}
\label{fig5}
\end{figure}
spatial profiles that are localized in space and endowed with an initial kick. One example of this is using the profile corresponding to the fundamental stationary mode. In all cases we have the generic behavior that mobility is enhanced for relatively wide profiles, and/or high values of the momentum $k$.

{\em Propagation control}. One of the major problems for achieving controllable steering of discrete solitons is the existence of an effective periodic potential, known as the Peierls-Nabarro (PN) potential, that appears as a result of lattice discreteness. While in the continuous case, the presence of translational invariance favors soliton propagation, in a discrete system
a minimum impulse is needed to effect soliton motion. 
The magnitude of the PN potential can be roughly estimated as $A^4$, where $A$ is the soliton amplitude\cite{PN}. We can shed some light into this problem by the use of strongly localized modes (SLMs)\cite{SLM}. We consider a stationary an odd SLM in the form
\be
\phi_{n}=\{ 0,0,\cdots,\epsilon \exp(-i k), 1, \epsilon \exp(i k),0,\cdots,0\} \phi_{o}
\ee
where $\epsilon \ll 1$. This mode has an associated Hamiltonian given by
\begin{eqnarray}
H_{odd}&=& -2 V \phi_{0}^2 (\epsilon+\epsilon^{*}) \cos(k) - \chi \phi_{o}^4 (1+2|\epsilon|^2+2 |\epsilon|^4)\nonumber\\
& & \approx -\chi \phi_{o}^4.
\end{eqnarray}
On the other hand, an even SLM has the form
\be
\phi_{n}=\{ 0,0,\cdots,\epsilon \exp(-i k), 1, \exp(i k),\epsilon \exp(2 i k),0,\cdots,0\} \tilde{\phi_{o}}
\ee
and an associated Hamiltonian
\be
H_{even} \approx
 -4 V \cos(k) \tilde{\phi_{0}}^2-3 \chi \tilde{\phi_{0}}^4
 \ee
On the other hand, the power content of these localized modes is given by
\be
P_{odd}\approx \phi_{0}^2 + O(\epsilon^2)
\ee
and
\be
P_{even} \approx 2 \tilde{\phi_{0}}^2 + O(\epsilon^2)
\ee
Now we assume that the odd and even SLM are different states of the {\em same} soliton. This implies that both SLMs possess the same norm (power). Therefore, $\phi_{0}^2 \approx 2 \tilde{\phi_{o}^2}$. The even Hamiltonian becomes
\be
H_{even} \approx
 -2 V \cos(k) \phi_{0}^2-(3/4) \chi \phi_{0}^4.
 \ee
The dynamical barrier can now be defined as the difference $\Delta=H_{odd}-H_{even}$, that is,
\be \Delta \approx 2 V \phi_{0}^2 \cos(k) - (1/4) \chi \phi_{0}^4.
\ee
We see that, to a first approximation, the barrier could be tuned by an appropriate choice of the amplitude, momentum and nonlinearity parameter. For the ideal case $\Delta=0$ the discrete soliton would propagate unimpeded across the lattice. If our objective is not to effect a free propagation, but to deliver the soliton at a given location (where it will remain due to selftrapping), like in a multiport switching, one could in principle, resort to an engineering of the couplings\cite{couplings} to bring the soliton from a given position to any desired site.

{\em Modulational Stability}. When dealing with the  dynamical evolution of the mDLNS equation, it is natural to ask under which circumstance the system will create discrete solitons instead of radiation. A clue about this comes from examining the linear stability of an initially uniform nonlinear profile. When this profile becomes unstable, the profile will tend to fragment and the largest fragments could serve as seeds for discrete solitons. Usually this depends on the strength of nonlinearity. Let us consider a solution of the form $C_{n}(t) = \phi \exp(i \lambda t)$. After inserting this solution into Eq.(\ref{one}), we conclude $\lambda = 2 V + 4 \chi \phi^2$ and therefore $C_{n}(t) = \phi\ \exp[i (2 V + 4 \phi^2)t]$.  We insert this solution into Eqs.(\ref{eq5}) and (\ref{eq6}) and obtain
\begin{eqnarray}
{\bf A}_{n m}& = & -2 V \delta_{n m} + V (\delta_{n,m}+  \delta_{n,m-1})\nonumber\\
{\bf B}_{n m}&=& (2V-4 
\chi \phi^2)\delta_{n m}-(V+2\chi \phi^2)(\delta_{n,m+1}+\delta_{n,m-1})\nonumber\\ 
\end{eqnarray}
\begin{figure}[t]
\noindent
\includegraphics[scale=0.6,angle=0]{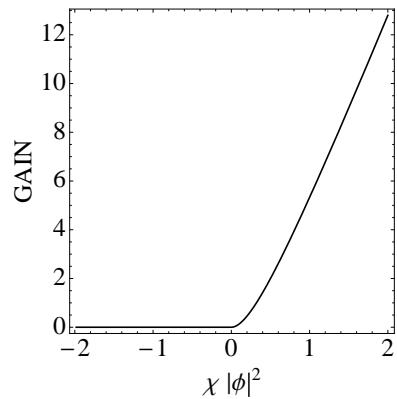}
\caption{Instability gain versus nonlinearity strength ($V=1$).}\label{fig6}
\end{figure}
and proceed with an analysis of the gain parameter. This procedure gives the same information as a semi-analytical method\cite{kivshar}. Results are  shown in Fig.6 which shows the instability gain versus the nonlinearity strength. For positive nonlinearity parameter the gain is positive signaling instability of the uniform profile and thus, adequate conditions for the creation of discrete solitons. For negative nonlinearity strength, the gain is identically zero.
This results are in agreement with those obtained for the DNLS in the limit of uniform initial profile\cite{kivshar}.

However, since in our case the nonlinearity parameter is proportional to the square of the electron-phonon  interaction, it is always positive and we can conclude that the system is modulationally unstable and thus, prone to generating discrete solitons. 

{\em Transport}. Finally, let us look at the transport properties of the mDNLS system. The typical thing to do is to examine the mean square displacement of an initially localized initial condition, at long evolution times\be \sigma^2 = {\sum_{n}{n^2 |C_{n}(t)|^2}\over{\sum_{n}|C_{n}(t)|^2}} \sim t^{\alpha}.
\ee
where, $C_{n}(0)=\delta_{n,0}$. For $\alpha=2$ we have ballistic motion, for $\alpha =1$ we have diffusive motion, for $1<\alpha<2$ we have super-diffusive motion, and for $0<\alpha<1$ we have sub-diffusive motion. 
\begin{figure}[t]
\noindent
\includegraphics[scale=0.7,angle=0]{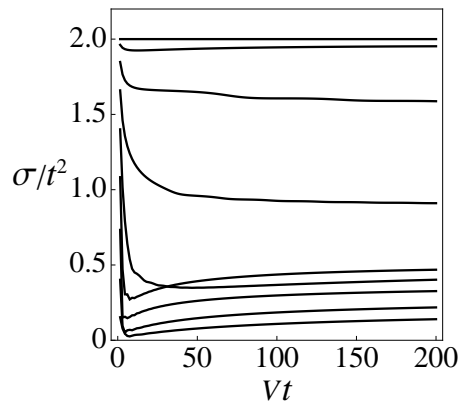}
\caption{Mean square displacement versus time for several nonlinearity parameter values. From top to bottom: $\chi=0,0.5,1,1.5,2,2.5,3,3.5,4$.}
\label{fig7}
\end{figure}
Figure 7 shows the evolution of the mean square displacement (MSD) in time for several values of the nonlinearity parameter. We have used a chain with $N=900$ which is sufficiently long to avoid reflection from the boundaries of a ballistic pulse for $z=200$. 
As we can see, after a transient, all the curves show an asymptotic exponent very close to $2$. This can be explained noting that as time evolves, the profile expands and brings the nonlinearity terms down. In other words, at long times the nonlinearity contribution in Eq.(\ref{one}) is negligible and the evolution becomes 
ballistic. Another thing we notice is how $\sigma$ (not the exponent) decreases with an increase in nonlinearity. This can be explained as follows: As $\chi$ increases, partial selftrapping of the excitation around the initial position increases as well and renormalizes the amount of radiation that can escape to infinity. 

\section{Conclusions}

We have studied the nonlinear bulk and surface modes of the modified discrete nonlinear Schr\"{o}dinger (mDNLS) equation. We have computed the linear stability of the lowest-order  modes and have also computed the modulational stability of the uniform solution. We conclude that the fundamental bulk mode is stable with a stability curve extending all the way from the high nonlinearity region down to the linear band. In general, higher modes need a minimum nonlinearity to exist and can posses alternate stability as a function of power content. The surface modes, on the other hand, all need a minimum nonlinearity threshold to exist, with a stable fundamental mode. We have also estimated the dynamical barrier for the motion of a localized excitation across the lattice and obtained an approximate expression in terms of the amplitude, initial momentum and nonlinearity. The modulation stability of the special uniform solution was computed, concluding that the system is modulationally unstable. This means that the system favors the creation of nonlinear localized excitations (solitons). Finally, we computed the asymptotic transport exponent, by examining the mean square displacement of an initially localized excitation. We found that, at long times, and as a result of norm conservation, nonlinear effects becomes smaller and smaller and, as a result the propagation exponent becomes the ballistic one in the limit of an infinite time.

It should be mentioned that similar results have been found for the complementary DNLS case. This is interesting and points to the robustness of the phenomenology found here concerning the modes stability, the discrete soliton propagation, and the asymptotic propagation exponent. This is encouraging in areas such as energy transport in biomolecules, where robust energy propagation mechanisms should be at work.

\acknowledgments
This work was supported in part by Fondecyt Grant   1160177 and Programa ICM grant RC130001.


\begin{thebibliography}{99}

\bibitem{davydov}
A. S. Davydov, {\em Solitons in Molecular Systems} (Kluwer Academic Publishers, Dordrecht, 1991); A. C. Scott, Phys. Rep.{\bf 217}, 1 (1992).

\bibitem{biomolecules}
{\em Davydov's Soliton Revisited: Self-trapping of Vibrational Energy in Protein}, P. L. Christensen and A. C. Scott, eds. (Plenum Press, New York, 1990 ).

\bibitem{prb51}
G. Kopidakis, C. M. Soukoulis and E. N. Economou, Phys. Rev. B {\bf 51}, 15038 (1995).

\bibitem{kopidakis}
G. Kopidakis, C. M. Soukoulis and E. N. Economou, Phys. Rev. B {\bf 49}, 7036 (1994).

\bibitem{tsironis}
G. Kalosakas, G. P. Tsironis and E. N. Economou, J. Phys.:Condens. Matter {\bf 6}, 7847 (1994).

\bibitem{dnls}
Panayotis G. Kevrekidis, {\em The Discrete Nonlinear Schr\"{o}dinger Equation} (Springer, Berlin Heidelberg 2009);  J. C. Eilbeck, P. S. Lomdahl, A. C. Scott
Physica D: Nonlinear Phenomena {\bf 16}, 318 (1985), M. I. Molina y G. P. Tsironis, Phys. Rev. Lett. {\bf 73}, 464 (1994); {\em The discrete nonlinear Schr\"{o}dinger equation-20 years on}, J. C. Eilbeck, M. Johansson,
Proceedings of the Conference on Localization and Energy Transfer in Nonlinear Systems, Madrid, Spain (2002) (World Scientific, 2003).

\bibitem{PN}
Yuri S. Kivshar and David K. Campbell, Phys. Rev. E {\bf 48}, 3077 (1993).

\bibitem{SLM}
Yuriy A. Kosevich, Phys. Rev. B {\bf 47}, 3138 (1993);
S. Darmanyan, A. Kobyakov, and F. Lederer
Phys. Rev. E {\bf 57}, 2344 (1998); S. Darmanyan, A. Kobyakov, E. Schmidt, and F. Lederer
Phys. Rev. E {\bf 57}, 3520 (1998).
 
\bibitem{couplings}
R. A. Vicencio, M. I. Molina and Y. S. Kivshar, Opt. Lett. {\bf 28}, 1942 (2003); R. A. Vicencio, M. I. Molina and Y. S. Kivshar, Phys. Rev. E. {\bf 70}, 026602 (2004).

\bibitem{kivshar}
Yuri S. Kivshar and Michel Peyrard, Phys. Rev. A {\bf 46}, 3198 (1992).



\end{thebibliography}
\end{document}